\newif\ifws
\ifws

\documentclass[entropy,article,submit,pdftex,oneauthor]{Definitions/mdpi}
\firstpage{1}
\pubvolume{1}
\issuenum{1}
\articlenumber{0}
\pubyear{2025}
\copyrightyear{2025}
\datereceived{ }
\daterevised{ } 
\dateaccepted{ }
\datepublished{ }
\hreflink{https://doi.org/} 

\Title{Chromatic Quantum Contextuality}

\TitleCitation{Chromatic Quantum Contextuality}

\Author{Karl Svozil $^{1,\dagger,\ddagger}$\orcidA{}}

\AuthorNames{Karl Svozil}

\isAPAStyle{%
       \AuthorCitation{Svozil, K.}
         }{%
        \isChicagoStyle{%
        \AuthorCitation{Svozil,Karl}
        }{
        \AuthorCitation{Svozil, K.}
        }
}

\address{%
$^{1}$ \quad Institute for Theoretical Physics,
TU Wien,
Wiedner Hauptstrasse 8-10/136,
1040 Vienna,  Austria; karl.svozil@tuwien.ac.at}

\corres{Correspondence: karl.svozil@tuwien.ac.at}

\abstract{
Chromatic quantum contextuality is a criterion of quantum nonclassicality based on (hyper)graph coloring constraints. If a quantum hypergraph requires more colors than the number of outcomes per maximal observable (context), it lacks a classical realization with $n$-uniform outcomes per context. Consequently, it cannot represent a ``completable'' non-contextual set of coexisting $n$-ary outcomes per maximal observable. This result serves as a chromatic analogue of the Kochen-Specker theorem. We present an explicit example of a four-colorable quantum logic in dimension three. Furthermore, chromatic contextuality suggests a novel restriction on classical truth values, thereby excluding two-valued measures that cannot be extended to $n$-ary colorings. Using this framework, we establish new bounds for the house, pentagon, and pentagram hypergraphs, refining previous constraints.
}

\keyword{contextuality; logic; hypergraph; chromatic number}

\usepackage{tikz}
\usetikzlibrary{decorations.markings}
\usetikzlibrary{calc}

\begin{document}


\else

\documentclass[%
 superscriptaddress,
  reprint,
 showpacs,
 showkeys,
 nofootinbib,
  amsmath,amssymb,
 pra,
  longbibliography,
  floatfix,
 ]{revtex4-2}

\usepackage[normalem]{ulem}

\usepackage{adjustbox}

\usepackage{hyperref}
\usepackage{amsmath}
\usepackage{amssymb}
\usepackage{amsthm}
\usepackage{bm} 
\usepackage{graphicx}

\RequirePackage{times}
\RequirePackage{mathptm}

\usepackage{url}
\usepackage[x11names]{xcolor}
\usepackage{eepic}
\usepackage{tikz}
\usetikzlibrary{decorations.markings}
\usetikzlibrary{calc}
\usepackage {pgfplots}
\pgfplotsset {compat=1.8}
\usepackage{epstopdf}
\usepackage[normalem]{ulem}
\sloppy

\newtheorem{theorem}{Theorem}
\newtheorem{comment}{Comment}
\newtheorem{proposition}{Proposition}
\newtheorem{corollary}{Corollary}
\newtheorem{fact}{Fact}
\newtheorem{lemma}{Lemma}
\theoremstyle{definition}
\newtheorem{definition}{Definition}

\newcommand{\seq}[1]{\mathbf{#1}}
\newcommand{\floor}[1]{\left\lfloor #1 \right\rfloor}
\newcommand{\ceil}[1]{\left\lceil #1 \right\rceil}
\newcommand{\abs}[1]{\left\lvert#1\right\rvert}
\newcommand{\rest}[2]{#1\!\!\restriction_{#2}}
\newcommand{\reste}[2]{#1\restriction_{#2}}
\newcommand{\N}{\mathbb{N}}
\newcommand{\Z}{\mathbb{Z}}
\newcommand{\Q}{\mathbb{Q}}
\newcommand{\R}{\mathbb{R}}
\newcommand{\C}{\mathbb{C}}
\newcommand{\alphabet}{\{0,1\}}
\newcommand{\B}{B^*}
\newcommand{\BI}{B^\omega}
\newcommand{\x}{\mathbf{x}}
\newcommand{\dom}{\text{dom}}
\newcommand{\cl}{\text{cl}}
\newcommand{\dd}{\mathrm{d}}

\newcommand{\bra}[1]{\left< #1 \right|}
\newcommand{\ket}[1]{\left| #1 \right>}

\newcommand{\iprod}[2]{\langle #1 | #2 \rangle}
\newcommand{\mprod}[3]{\langle #1 | #2 | #3 \rangle}
\newcommand{\oprod}[2]{| #1 \rangle\langle #2 |}

\begin{document}

\title{Chromatic Quantum Contextuality}

\author{Karl Svozil}
\email{karl.svozil@tuwien.ac.at}
\homepage{http://tph.tuwien.ac.at/~svozil}

\affiliation{Institute for Theoretical Physics,
TU Wien,
Wiedner Hauptstrasse 8-10/136,
1040 Vienna,  Austria}

\date{\today}

\begin{abstract}
Chromatic quantum contextuality is a criterion of quantum nonclassicality based on (hyper)graph coloring constraints. If a quantum hypergraph requires more colors than the number of outcomes per maximal observable (context), it lacks a classical realization with $n$-uniform outcomes per context. Consequently, it cannot represent a ``completable'' non-contextual set of coexisting $n$-ary outcomes per maximal observable. This result serves as a chromatic analogue of the Kochen-Specker theorem. We present an explicit example of a four-colorable quantum logic in dimension three. Furthermore, chromatic contextuality suggests a novel restriction on classical truth values, thereby excluding two-valued measures that cannot be extended to $n$-ary colorings. Using this framework, we establish new bounds for the house, pentagon, and pentagram hypergraphs, refining previous constraints.
\end{abstract}

\keywords{contextuality, logic, hypergraph, chromatic number}

\maketitle

\fi

\section{Contexts as maximal observables}

In operator-valued arguments, a context can be seen as being capable of potentially encoding and utilizing
\textit{maximal knowledge} of a system---all that can be conceivably extracted from a single `maximal' measurement~\cite{zeil-99}.

A context can, through the spectral theorem, be identified with an orthonormal basis with elements $| e_i \rangle $,
or, alternatively, the associated mutually orthogonal, self-adjoint projection operators  $| e_i \rangle  \langle  e_i |$.
These can be bundled together by a non-degenerate spectral sum (decomposition) as
$
\sum_i a_i | e_i \rangle  \langle  e_i |
$
with real-valued $a_i$ which are mutually distinct, to form a self-adjoint \textit{maximal} operator
introduced by von Neumann~\cite[Satz~8, p.~221f]{v-neumann-31}; a good description is given by Halmos~\cite[\S~84, p.~171f]{halmos-vs}.

A maximal (and thus non-degenerate) operator, or its corresponding context,
represents the totality of what can be ideally measured---no more, no less.
In this case, the measurement \textit{resolution} is at its finest, corresponding to
individual basis elements \( | e_i \rangle \). Equivalently, it pertains to the (mutually orthogonal)
one-dimensional subspaces spanned by these basis elements, and the respective property
of ``being in the state \( | e_i \rangle \)''.

In contrast, a two-valued measure resolves a single one-dimensional subspace spanned by $| e_i \rangle $,
assigning it the value $1$,
while the rest, an $(n-1)$-dimensional subspace in an $n$-dimensional Hilbert space, is assigned the value $0$.
Any orthonormal basis element of this $(n-1)$-dimensional subspace is assigned the value $0$.

Therefore, a \textit{coloring} by mutually different numbers, parameters, outcomes or colors $a_i$ yields not only a finer resolution than two-valued states can offer,
but corresponds to an optimal experimental extraction of data from a state by a maximal operator.
In physics, we need to insist on (at least in principle) maximal conceivable resolution---all
that could at least in principle be measured.

\section{Connection to chromatic number of hypergraphs}

Hypergraphs~\cite{Bretto-MR3077516} are extensively utilized in quantum logics to model propositions and the contexts (Boolean subalgebras) to which they belong~\cite{greechie:71,nav:91,Mckay2000,doi:10.1063/1.3491766}.
In this framework, each Boolean subalgebra---also known as a block, maximal operator, orthonormal basis, or context---is represented by a hyperedge in the hypergraph.
These hyperedges are visually depicted as smooth lines connecting the vertices that correspond to the propositions within that subalgebra.
The structure of the hypergraph captures the orthogonality relations among propositions, where propositions within the same hyperedge are pairwise orthogonal~\cite{lovasz-79,Portillo-2015}.
For a recent detailed exposition of these concepts, including precise definitions and illustrative examples, the reader is referred to Ref.~\cite{svozil-2021-chroma}.

In the context of hypergraph coloring each maximal observable corresponds to a hyperedge,
and the vertices represent the possible outcomes or states associated with that observable.
The \textit{exclusivity} requirement---that no hyperedge can have two or more of its vertices
colored the same---reflects the quantum mechanical constraint that a maximal observable
cannot yield two or more outcomes for any of its possible eigenstates.

A further \textit{completeness} requirement states that all colors must occur in each hyperedge,
meaning that no hyperedge lacks a color necessary for coloring the entire hypergraph.
Thus, the chromatic number of the hypergraph provides a measure of the minimal number
of distinct outcomes needed to satisfy these constraints.

A coloring of a hypergraph is said to be \textit{admissible} if it is both exclusive and complete.
If no admissible coloring exists---that is, if the chromatic number exceeds the number of vertices per hyperedge---this
indicates that no conceivable and possible coloring (relative to admissibility, that is, the exclusivity
and completeness rules) exists. This `excess of required outcomes', in turn, highlights the nonclassical nature of
the respective collection of quantum observables corresponding to the hypergraph
(with vector vertex labels).

In what follows, we shall only consider \textit{$n$-uniform} hypergraphs with an equal number $n$ of vertices per edge.
As argued earlier, every edge of such a hypergraph can be identified with a context, and a maximal observable.
We shall consider colorings of such hypergraphs as color assignments to its vertices
such that no hyperedge has vertices with the same color: every hyperedge contains vertices in $n$ different colors.

The chromatic number \( k \) of a hypergraph is the \textit{minimal} number of colors required to achieve a coloring
satisfying exclusivity. Note that this not necessarily means that the coloring satisfies completeness.
We note in passing that, unlike the terminology used here, a proper coloring of a hypergraph often refers to an assignment of colors
to its vertices such that each hyperedge contains at least two vertices of different colors.

Moreover, a \( k \)-uniform proper coloring of a hypergraph is an assignment of colors to its vertices
such that all colors are assigned to an equal number of vertices~\cite{berlov-2014}.
This implies that the assignment of colors partitions the set of vertices into disjoint subsets of equal size.

If $k=n$ we obtain a `canonical' $k$-uniform proper coloring of $n$-uniform hypergraphs,
which are particularly important for physics:
Any such coloring can be reduced to a two-valued state through \textit{aggregation}:
assigning a single color the value $1$ while mapping all other $k-1$ colors to $0$~\cite{meyer:99}.
However, the inverse is not possible:
As will be discussed later, the mere existence of two-valued states does not guarantee a chromatic number $n$~\cite[Appendix~B, p.~032104-16f]{svozil-2021-chroma}.

Any admissible coloring---respecting both exclusivity and completeness---corresponds to a value assignment that maintains maximal resolution within each context. In the hypergraph coloring framework, each maximal observable is represented by a hyperedge, with vertices corresponding to its possible outcomes or eigenstates. The constraint that no hyperedge can contain two or more identically colored vertices reflects the quantum mechanical principle that a maximal observable cannot yield identical outcomes for distinct eigenstates. Additionally, each hyperedge must include all necessary colors to properly color the hypergraph. Since the chromatic number of the hypergraph quantifies the exact number of distinct outcomes required to meet these constraints,
any deviation exceeding the uniform number of vertices per hyperedge indicates the nonclassical nature of quantum systems.

If we can identify a quantum-representable hypergraph---one that permits a faithful orthogonal
representation~\cite{lovasz-79,Cabello-2010-ncoptaa,Portillo-2015}---that is uniform with $n$ vertices per hyperedge and
has a chromatic number exceeding $n$, then we can demonstrate that this configuration does not correspond
to a physically realizable (classical counterfactual)
measurement setup with noncontextual, coexisting uniform outcomes.
We may perceive this as a form of \textit{chromatic contextuality}.

Chromatic contextuality differs from the theorems of Kochen-Specker,
Greenberger-Horne-Zeilinger~\cite{svozil-2024-convert-pra-externalfigures}, or Hardy's paradox~\cite{svozil-2020-hardy},
which can all be expressed as arguments involving two-valued states, although some of them can be written in terms of
operator values.
The mere existence of a two-valued state---amounting to, within a given context, assigning a unique value (say, `1') to one outcome and a different, single value (say, `0') to all other outcomes---is, by itself, insufficient to prove that mutually distinct outcomes---associated with that context's maximal observable---pre-exist;
in particular, when demanding uniform consistency across all possible intertwining contexts (thereby necessarily involving counterfactual nondegenerate outcomes also for unchosen measurements).
This is true even if the set of two-valued states is separable---meaning that there exists at least one two-valued state that separates every pair of vertices---because
such configurations may not allow the potential acquisition (through counterfactual experiments) of maximal knowledge (per context).

Clearly, in the Kochen-Specker cases, there is no two-valued state and thus no coloring.
However, we might hope to find find `smaller hypergraphs' (with a `small' number of edges or vertices)
that have no coloring with $n$ colors but still have (even a separating set of) two-valued states conforming to the demarcation Theorem~0 of Kochen and Specker~\cite{kochen1}.

\section{Previous results}

There exists a set representable hypergraph that does not allow a coloring whose required minimal number of colors
exceeds the number of vertices per edge:
Its graph $G_{32}$ has been early discussed by Greechie~\cite[Figure~6, p.~121]{greechie:71}
        (see also Refs.~\cite{Holland1975,Bennett-MC-1970,Greechie1974,Greechie-Suppes1976}).
$G_{32}$ is a 3-uniform hypergraph with 15 bi-intertwined vertices in 10 contexts.
It supports a separating set of two-valued states.
Its chromatic number is 4~\cite[Appendix~B, p.~032104-16f]{svozil-2021-chroma}.

\section{Chromatic number of the Yo-Oh hypergraph}

\begin{figure*}
\begin{center}
\begin{tabular}{cc}
		\begin{tikzpicture}  [scale=0.3]

\newdimen\ms
\ms=0.1cm

\tikzstyle{every path}=[line width=1pt]
\tikzstyle{c3}=[circle,inner sep={\ms/8},minimum size=3*\ms]
\tikzstyle{c2}=[circle,inner sep={\ms/8},minimum size=2*\ms]
\tikzstyle{c1}=[circle,inner sep={\ms/8},minimum size=1.1*\ms]

		\path
     (-2.50, 6.87    ) coordinate(2)     
			  (-6.33,   3.65  ) coordinate(4)    
			  (-7.20, -1.27   ) coordinate(6)       
			  (-4.70, -5.60   ) coordinate(8)       
			  (0, -7.31       ) coordinate(10)       
			  (4.70, -5.60    ) coordinate(12)        
     (7.20, -1.27    ) coordinate(14)       
			  (6.33, 3.65     ) coordinate(16)       
			  (2.50, 6.87     ) coordinate(18)    

			  (0, 6.87        ) coordinate(1)     
			  (-4.42, 5.26    ) coordinate(3)
			  (-6.76, 1.19    ) coordinate(5)
			  (-5.95, -3.43   ) coordinate(7)     
			  (-2.35, -6.45   ) coordinate(9)
     (2.35, -6.45    ) coordinate(11)
			  (5.95,-3.43     ) coordinate(13)    
			  (6.76, 1.19     ) coordinate(15)
		   (4.42, 5.26     ) coordinate(17)

     (0,0) coordinate(19)

(0, 10.3       ) coordinate(20)     
			  (-8.7, -5.2    ) coordinate(21)     
			  (8.7,-5.2    ) coordinate(22)    

     (-1.25, 3.435 ) coordinate(23)
     (3.60, -0.635 ) coordinate(24)
     (-2.35, -2.80 ) coordinate(25)
     (1.9425, -3.2875) coordinate(26)
     (1.875, 3.325) coordinate(27)
     (-3.8175, -0.04) coordinate(28)
;


		
\draw [rounded corners=20mm,color=lightgray]     (20) -- (21) -- (22) --  cycle;

\draw [color=green] (18) -- (1) -- (2);
\draw [color=violet] (2) -- (3) -- (4);
\draw [color=gray] (4) -- (5) -- (6);
\draw [color=magenta] (6) -- (7) -- (8);
\draw [color=blue] (8) -- (9) -- (10);
\draw [color=red] (10) -- (11) -- (12) ;

\draw [color=lime] (12) -- (13) -- (14);
\draw [color=cyan] (14) -- (15) -- (16);
\draw [color=orange] (16) -- (17) -- (18);

\draw [color=teal] (19) -- (2);
\draw [color=olive] (19) -- (8);
\draw [color=pink] (19) -- (14);

\draw [color=brown] (4) -- (12);
\draw [color=black] (10) -- (18);
\draw [color=darkgray] (16) -- (6);


\draw (19) coordinate[c3,fill=teal];
\draw (19) coordinate[c2,fill=olive];
\draw (19) coordinate[c1,fill=pink,label={[xshift=1.5pt]180:{\small $h_0$}}];

\draw (1)  coordinate[c3,fill=lightgray,label=90:$z_1$];
\draw (1)  coordinate[c2,fill=green];

\draw (2)  coordinate[c3,fill=teal,label=90:$y_1^-$];

\draw (3)  coordinate[c2,fill=violet];

\draw (4)  coordinate[c3,fill=gray,label=180:$h_1$];
\draw (4)  coordinate[c2,fill=brown];
\draw (4)  coordinate[c1,fill=violet];

\draw (5)  coordinate[c2,fill=gray,label=0:$ $];

\draw (6)  coordinate[c3,fill=darkgray,fill=gray,label=180:$y_2^+$];
\draw (6)  coordinate[c2,fill=magenta];
\draw (6)  coordinate[c1,fill=gray];

\draw (7)  coordinate[c3,fill=lightgray,label=180:$z_2$];
\draw (7)  coordinate[c2,fill=magenta];

\draw (8)  coordinate[c3,fill=olive,label=180:$y_2^-$];
\draw (8)  coordinate[c2,fill=blue];
\draw (8)  coordinate[c1,fill=magenta];

\draw (9)  coordinate[c2,fill=blue,label=0:$ $];

\draw (10) coordinate[c3,fill=blue,label=270:$h_2$];
\draw (10) coordinate[c2,fill=red];
\draw (10) coordinate[c1,fill=black];

\draw (11) coordinate[c2,fill=red,label=0:$ $];

\draw (12) coordinate[c3,fill=brown,label=0:$y_3^+$];
\draw (12) coordinate[c2,fill=red];
\draw (12) coordinate[c1,fill=lime];

\draw (13) coordinate[c3,fill=lightgray,label=0:$z_3$];
\draw (13) coordinate[c2,fill=lime];

\draw (14) coordinate[c3,fill=cyan,label=0:$y_3^-$];
\draw (14) coordinate[c2,fill=pink];
\draw (14) coordinate[c1,fill=lime];

\draw (15) coordinate[c2,fill=cyan,label=0:$ $];

\draw (16) coordinate[c3,fill=darkgray,label=0:$h_3$];
\draw (16) coordinate[c2,fill=orange];
\draw (16) coordinate[c1,fill=cyan];

\draw (17) coordinate[c2,fill=orange,label=0:$ $];

\draw (18) coordinate[c3,fill=orange,label=90:$y_1^+$];
\draw (18) coordinate[c2,fill=black];
\draw (18) coordinate[c1,fill=green];

\draw (23) coordinate[c2,fill=teal];
\draw (24) coordinate[c2,fill=pink];
\draw (25) coordinate[c2,fill=olive];
\draw (26) coordinate[c2,fill=brown];
\draw (27) coordinate[c2,fill=black];
\draw (28) coordinate[c2,fill=darkgray];

		\end{tikzpicture}
&
\begin{tikzpicture}  [scale=0.4]
\newdimen\ms
\ms=0.1cm

\tikzstyle{every path}=[line width=1pt]
\tikzstyle{c3}=[circle,inner sep={\ms/8},minimum size=3*\ms]
\tikzstyle{c2}=[circle,inner sep={\ms/8},minimum size=2*\ms]
\tikzstyle{c1}=[circle,inner sep={\ms/8},minimum size=1.1*\ms]

\newdimen\R
\R=6cm
\newdimen\K
\K=3cm

\path
  ({90 + 0 * 360 /5}:\R      ) coordinate(1)
  ({90 + 360 /10 + 0 * 360/5} : {\R * 0.6881/0.8507} ) coordinate(2)
  ({90 + 1 * 360 /5}:\R   ) coordinate(3)
  ({90 + 360 /10 + 1 * 360/5} : {\R * 0.6881/0.8507} ) coordinate(4)
  ({90 + 2 * 360 /5}:\R  ) coordinate(5)
  ({90 + 360 /10 + 2 * 360/5} : {\R * 0.6881/0.8507} ) coordinate(6)
  ({90 + 3 * 360 /5}:\R  ) coordinate(7)
  ({90 + 360 /10 + 3 * 360/5} : {\R * 0.6881/0.8507} ) coordinate(8)
  ({90 + 4 * 360 /5}:\R     ) coordinate(9)
  ({90 + 360 /10 + 4 * 360/5} : {\R * 0.6881/0.8507} ) coordinate(10)
  ({90 + 0 * 360 /5}:\K      ) coordinate(11)
  ({90 + 360 /10 + 0 * 360/5} : {\K * 0.6881/0.8507} ) coordinate(12)
  ({90 + 1 * 360 /5}:\K   ) coordinate(13)
  ({90 + 360 /10 + 1 * 360/5} : {\K * 0.6881/0.8507} ) coordinate(14)
  ({90 + 2 * 360 /5}:\K  ) coordinate(15)
  ({90 + 360 /10 + 2 * 360/5} : {\K * 0.6881/0.8507} ) coordinate(16)
  ({90 + 3 * 360 /5}:\K  ) coordinate(17)
  ({90 + 360 /10 + 3 * 360/5} : {\R * 0.6881/0.8507} ) coordinate(18)
  ({90 + 4 * 360 /5}:\K     ) coordinate(19)
  ({90 + 360 /10 + 4 * 360/5} : {\K * 0.6881/0.8507} ) coordinate(20)

  ({90 + 2 * 360 /5}:{(\R+\K)/2}) coordinate(21)

;


\draw [color=orange] (1) -- (2) -- (3);
\draw [color=darkgray] (3) -- (4) -- (5);
\draw [color=gray] (5) -- (6) -- (7);
\draw [color=violet] (7) -- (8) -- (9);
\draw [color=green] (9) -- (10) -- (1);    %

\draw [color=blue] (11) -- (15) ;
\draw [color=olive] (15) -- (19);
\draw [color=pink] (19) -- (13);
\draw [color=lime] (13) -- (17);
\draw [color=red] (17) -- (11);

\draw ($ (11) !.5! (15) $) coordinate[c2,fill=blue];  %
\draw ($ (15) !.5! (19) $) coordinate[c2,fill=olive];  %
\draw ($ (19) !.47! (13) $) coordinate[c2,fill=pink];  %
\draw ($ (17) !.5! (11) $) coordinate[c2,fill=red];  %

\draw [color=black] (1) -- (11);
\draw [color=cyan] (3) -- (13);
\draw [color=magenta] (5) -- (15);
\draw [color=brown] (7) -- (17);
\draw [color=teal] (9) -- (19);

\draw ($ (1) !.5! (11) $) coordinate[c2,fill=black];  %
\draw ($ (3) !.5! (13) $) coordinate[c2,fill=cyan];  %
\draw ($ (7) !.5! (17) $) coordinate[c2,fill=brown];  %
\draw ($ (9) !.5! (19) $) coordinate[c2,fill=teal];  %

\draw [color=lightgray] (10)  to   [out=-140,in=100] ($ (13) !.5! (17) $)  to   [out=-90,in=0] (21);
%
%
\draw (1) coordinate[c3,fill=black,label=90:$y_1^+$];  %
\draw (1) coordinate[c2,fill=green];   %
\draw (1) coordinate[c1,fill=orange];  %
\draw (2) coordinate[c2,fill=orange];    %
\draw (3) coordinate[c3,fill=cyan,label=180:$h_3$];   %
\draw (3) coordinate[c2,fill=darkgray]; %
\draw (3) coordinate[c1,fill=orange];  %
\draw (4) coordinate[c2,fill=darkgray];  %
\draw (5) coordinate[c3,fill=magenta,label=270:$y_2^+$];  %
\draw (5) coordinate[c2,fill=gray];  %
\draw (5) coordinate[c1,fill=darkgray];  %
\draw (6) coordinate[c2,fill=gray];
\draw (7) coordinate[c3,fill=brown,label=270:$h_1$];  %
\draw (7) coordinate[c2,fill=violet];  %
\draw (7) coordinate[c1,fill=gray];  %
\draw (8) coordinate[c2,fill=violet];  %
\draw (9) coordinate[c3,fill=green,label=0:$y_1^-$];
\draw (10) coordinate[c2,fill=green,label=0:$z_1$];  %
\draw (10) coordinate[c1,fill=lightgray];  %
\draw (11) coordinate[c3,fill=black,label=180:$h_2$];  %
\draw (11) coordinate[c2,fill=red];  %
\draw (11) coordinate[c1,fill=blue]; %
\draw (13) coordinate[c3,fill=cyan,label=90:$y_3^-$]; %
\draw (13) coordinate[c2,fill=lime];  %
\draw (13) coordinate[c1,fill=pink];  %
\draw (15) coordinate[c3,fill=magenta,label=180:$y_2^-$]; %
\draw (15) coordinate[c2,fill=olive]; %
\draw (15) coordinate[c1,fill=blue]; %

\draw (17) coordinate[c3,fill=brown,label=0:$y_3^+$];  %
\draw (17) coordinate[c2,fill=red];   %
\draw (17) coordinate[c1,fill=lime]; %

\draw (19) coordinate[c3,fill=teal,label=90:$h_0$]; %
\draw (19) coordinate[c2,fill=pink]; %
\draw (19) coordinate[c1,fill=olive]; %

\draw (21) coordinate[c2,fill=magenta];  %
\draw (21) coordinate[c1,fill=lightgray,label=-15:$z_2$];  %

\draw ($ (13) !.5! (17) $) coordinate[c2,fill=lightgray];  %
\draw ($ (13) !.5! (17) $) coordinate[c1,fill=lime,label=45:$z_3$];  %

\end{tikzpicture}
\\
(a)&(b)
\end{tabular}
\end{center}
\caption{\label{2017-b-f-Yu-2012} (Color online) Redraw~\cite[Chapter~12, p.~92]{svozil-pac} of two equivalent representations (a) and (b) of a Petersen graph-like
(with one additional context connecting
$z_1$,
$z_2$, and
$z_3$)
hypergraph of the logic considered by Yu and Oh~\cite[Fig.~2]{Yu-2012}.
The set of two-valued states enforces at most one of the four atoms $h_0$, $h_1$, $h_2$, $h_3$ to be 1.
The logic has a (quantum) realization in $\mathbb{R}^3$
consisting of the 25 projections;  associated with the one dimensional subspaces
spanned by  the 13 vectors from the origin $\left(0,0,0\right)^\intercal$ to
$z_1 = \left( 1, 0, 0 \right)^\intercal $,
$z_2 = \left( 0, 1, 0 \right)^\intercal $,
$z_3 = \left( 0, 0, 1 \right)^\intercal $,
$y^-_1 = \left( 0, 1, -1 \right)^\intercal $,
$y^-_2 = \left( 1, 0, -1 \right)^\intercal $,
$y^-_3 = \left( 1, -1, 0 \right)^\intercal $,
$y^+_1 = \left( 0, 1, 1 \right)^\intercal $,
$y^+_2 = \left( 1, 0, 1 \right)^\intercal $,
$y^+_3 = \left( 1, 1, 0 \right)^\intercal $,
$h_0 = \left( 1, 1, 1 \right)^\intercal $,
$h_1 = \left( -1, 1, 1 \right)^\intercal $,
$h_2 = \left( 1, -1, 1 \right)^\intercal $,
$h_3 = \left( 1, 1, -1 \right)^\intercal $,
respectively~\cite{Yu-2012}.
}
\end{figure*}
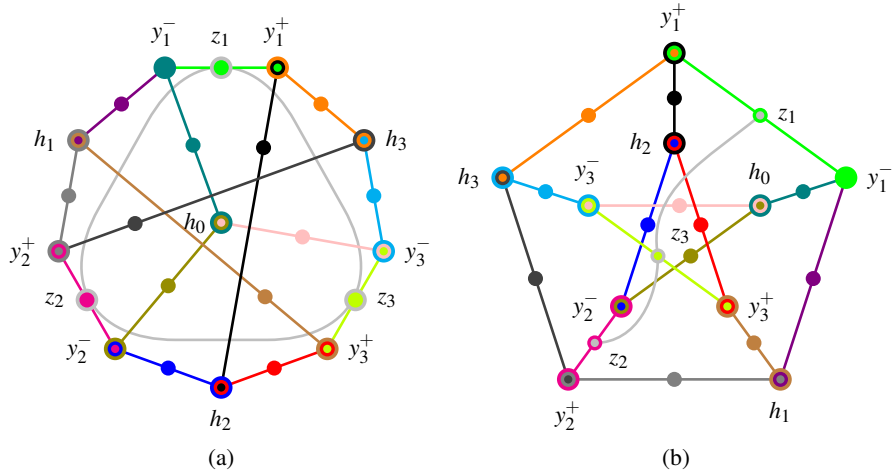

In a proof by contradiction, suppose the Yo-Oh configuration~\cite{Yu-2012} of quantum observables (propositions),
as depicted by a 3-uniform hypergraph in a previous publication~\cite[Chapter~12, p.~92]{svozil-pac} and redrawn in Figure~\ref{2017-b-f-Yu-2012},
can be colored with three colors: red, green, and blue.

Without loss of generality, we can assume $h_0$ is red.
Then, $y_1^-$, $y_2^-$, and $y_3^-$ must be either green or blue.

Assuming all are colored green leads to a contradiction, as the context $\{z_1,z_2,z_3\}$ would then lack the color green.

Therefore, at least one of $y_1^-$, $y_2^-$, and $y_3^-$ must be colored differently.
By symmetry, without loss of generality, let $y_1^-$ and $y_2^-$ be green, and $y_3^-$ be blue, as depicted in Figures~\ref{2025-b-f-Yu-2012-c1} and~\ref{2025-b-f-Yu-2012-c2}, (a) and (b), respectively.
Then, $z_1$ as well as $z_2$ cannot be green, and $z_3$ cannot be blue.

\subsection*{Case 1}

Suppose that $z_1$ is red.
Then, $z_2$ must be blue, and $z_3$ must be green, as depicted in Figure~\ref{2025-b-f-Yu-2012-c1}(c) and (d).
We can now assign colors to the three contexts
$\{
y_1^+ , y_1^- , z_1
\}$,
$\{
y_2^+ , y_2^- , z_2
\}$, and
$\{
y_3^+ , y_3^- , z_3
\}$
by assigning
blue to $y_1^+$,
red to $y_2^+$, and
red to $y_3^+$, respectively, as depicted in Figure~\ref{2025-b-f-Yu-2012-c1}(e) and (f).

Consequently, $h_2$ cannot be red because $y_3^+$ is red, nor green, as $y_2^-$ is green,
nor blue as $y_1^+$ is blue,
and all $y_3^+$, $y_2^-$, and $y_1^+$ are adjacent to $h_2$.

\subsection*{Case 2}

Suppose that $z_1$ is blue.
Then $z_2$ must be red, and $z_3$ must be green, as depicted in Figure~\ref{2025-b-f-Yu-2012-c2}(c) and (d).

Consequently, $y_1^+$ must be red, $y_2^+$ must be blue, and $y_3^+$ must be red.
Additionally, $h_3$ must be green, and $h_2$ must be blue.
As before, we can now assign colors to the three contexts
$\{
y_1^+ , y_1^- , z_1
\}$,
$\{
y_2^+ , y_2^- , z_2
\}$, and
$\{
y_3^+ , y_3^- , z_3
\}$
by assigning
red to $y_1^+$ and $y_3^+$, and
blue to $y_2^+$, respectively, as depicted in Figure~\ref{2025-b-f-Yu-2012-c2}(e) and (f)..

Now $h_1$ cannot be red because $y_3^+$ is red, nor green, as $y_1^-$ is green, nor blue as $y_2^+$ is blue,
and all $y_3^+$, $y_1^-$, and $y_2^+$ are adjacent to $h_1$.

It is not difficult to work out a coloring of the Yu-Oh hypergraph with four colors.
Therefore, its chromatic number is 4.
In passing, we note that it has a separating set of 24 two-valued measures.


\begin{figure*}
\begin{center}

\end{center}
\caption{\label{2025-b-f-Yu-2012-c1} (Color online)
Case 1 of the proof that the Yu-Oh hypergraph depicted in Figure~\ref{2017-b-f-Yu-2012} cannot be (noncontextually)
colored by three colors: its chromatic number is four.
}
\end{figure*}


\begin{figure*}
\begin{center}
%
\end{center}
\caption{\label{2025-b-f-Yu-2012-c2} (Color online)
Case 2 of the proof that the Yu-Oh hypergraph depicted in Figure~\ref{2017-b-f-Yu-2012} cannot be (noncontextually)
colored by three colors: its chromatic number is four.
}
\end{figure*}

\section{Summary}

Chromatic contextuality, characterized by the impossibility of admissibly coloring an $n$-uniform hypergraph with $n$ colors per hyperedge,
presents a compelling case against the existence of classical (noncontextual) hidden parameters.
In this way, chromatic contextuality resembles Kochen-Specker contextuality,
which is defined by the absence of uniform two-valued states~\cite{kochen1,Cabello-1996ega},
or admissible states~\cite{2015-AnalyticKS}.
Chromatic contextuality can be viewed as an extreme form of operator-valued argument,
given that the number of values involved is equivalent to the number of vertices in each context-edge of the hypergraph.

Chromatic contextuality, in contrast to Kochen-Specker contextuality, constitutes a distinct criterion for nonclassicality:
The respective hypergraphs and the collection of quantum observables they represent
may still support two-valued states and even permit (though not necessarily imply) classical embeddability
through a separating set of two-valued states.
This is exemplified by Greechie's \( G_{32} \) hypergraph,
which admits a set representation in terms of a partition logic~\cite[Appendix~B, p.~032104-16f]{svozil-2021-chroma}.

Moreover, any \( n \)-coloring can be directly converted into a two-valued state---indeed, into \( n \) two-valued states---by
aggregation, that is, by
reducing or folding the number of \( n \) colors into two.
This can be done by identifying a single color with the value \( 1 \) and all remaining colors with \( 0 \).
In this way, the set of observables encoding the hypergraph is equi-partitioned.

Alternatively, an \( n \)-coloring can be used by identifying more than one color with a non-zero value~\cite{ramanathan-2024},
for example, identifying two colors with the value \( 1/2 \).
A single coloring thus defines a canonical set of \( n \) two-valued states
covering the entire hypergraph~\cite[Appendix~A]{svozil-2021-chroma}.

However, the converse is not true: The existence of even separating sets of two-valued states does not imply the existence of a coloring,
as demonstrated by the aforementioned example of $G_{32}$.

It is quite remarkable that not all such colorings can be derived from the nonexclusive hypergraph coloring scheme discussed above.
In particular, the exotic two-times-\( 1/2 \) coloring of the pentagon (or house or pentagram) on intertwining context-hyperedges,
as exposed by Greechie~\cite[Figure~5, p.~186]{greechie-1974} and Wright~\cite[$\omega_{0}$, p.~268]{wright:pent},
cannot be obtained through identifying colors. The reason for this is a parity argument:
For an odd number of hyperedges, such as five, there cannot exist a coloring with the same color assigned to all (odd-numbered)
intertwining vertices.

By the same parity reasoning, one of the eleven two-valued states of the pentagon,
in which all the values are assigned to the nonintertwining `middle' vertices, does not originate from the aggregation
from a hypergraph coloring. Suppose one of the colors, say red, is always centered in the middle;
this would imply that the other two colors, say green and blue, must alternate at the five vertices with intertwining contexts.
However, for an odd number of such intertwining vertices, this leads to a disallowed configuration;
more explicitly, to green-red-blue-red-green-red-blue-red-\textit{green-red-green} (from cycle).
If this two-valued state is eliminated, the Hull computation still yields Klyachko's pentagram inequality~\cite[Equation~(5)]{Klyachko-2008},
along with an additional upper bound:
$1 \ge A_{13}+A_{35}+A_{57}+A_{79}+A_{91}\ge - 3$.
This explicit house-pentagon-pentagram example also demonstrates that, for hyperedges with more than two vertices,
not all two-valued states can necessarily be derived by aggregation---the
irreversible mapping of one color to the value $1$ and all other colors to $0$.
The Bub and Stairs inequality~\cite{Bub-2009} remains unaffected as they do not use the `color-fobidden middle-center' two-valued state.
This topic is too broad to be fully addressed within the scope of this paper.

For physical reasons we suggest that every two-valued state should ultimately originate from `perfect',
that is, maximal, measurements corresponding to colorings.
Therefore, only two-valued states that are derived through aggregation should be considered when deriving,
for instance, Boole-Bell-type inequalities by solving the Hull problem for the respective correlation polytope~\cite{froissart-81,pitowsky-86}.
Two-valued states that cannot be represented by aggregation from colorings are unphysical in the sense that no
(classical) maximal observables or contexts exist that could justify their inclusion.

The chromatic analogue of True-Implies-False (TIFs) and True-Implies-True (TITs) gadgets~\cite{2018-minimalYIYS}, which exhibit a distinct form of contextuality,
such as Hardy-type~\cite{svozil-2020-hardy}, has not yet been explored in the context of hypergraph colorings.
Additionally, the concept of chromatic separability, similar to Kochen-Specker's demarcation criterion of (non)separability with respect to two-valued states, remains to be discussed.
For instance, does nonseparability by two-valued states imply chromatic nonseparability?
We note that Kochen and Specker's $\Gamma_3$~\cite{kochen1} is also color nonseparable.
Is aggregation sufficient  for the inheritance of such properties?
These topics require future investigation.

\ifws
\vspace{6pt}

\funding{
This research was funded in whole or in part by the Austrian Science Fund (FWF) [Grant DOI:10.55776/I4579].
}

\acknowledgments{
The author gratefully acknowledges discussions with Mohammad Hadi Shekarriz.

Josef Tkadlec has kindly provided a program to find all two-valued states and important properties thereof, such as (non)separability,
given the set of contexts of a logic.

}

\conflictsofinterest{The author declares no conflicts of interest.
The funders had no role in the design of the study; in the collection, analyses, or interpretation of data;
in the writing of the manuscript; or in the decision to publish the results.}




\begin{adjustwidth}{-\extralength}{0cm}

\reftitle{References}


 \bibliography{svozil}

\PublishersNote{}
\end{adjustwidth}


\else

\begin{acknowledgments}
The author gratefully acknowledges discussions with Mohammad Hadi Shekarriz.

Josef Tkadlec has kindly provided a program to find all two-valued states and important properties thereof, such as (non)separability,
given the set of contexts of a logic.

This research was funded in whole or in part by the Austrian Science Fund (FWF) [Grant DOI:10.55776/I4579].
\end{acknowledgments}

\bibliographystyle{apsrev}
\bibliography{svozil}

\fi

\end{document}